# Of Mites and Men: Johannes Kepler on Stars and Size
(with an English translation of Chapter 16 of his 1606 *De Stella Nova*)

Christopher M. Graney
Jefferson Community & Technical College
Louisville, KY 40272
christopher.graney@kctcs.edu

ABSTRACT: In his 1606 *De Stella Nova*, Johannes Kepler attempted to answer Tycho Brahe's argument that the Copernican heliocentric hypothesis required all the fixed stars to dwarf the Sun, something Brahe found to be a great drawback of that hypothesis. This paper includes a translation into English of Chapter 16 of *De Stella Nova*, in which Kepler discusses this argument, along with brief outlines of both Tycho's argument and Kepler's answer (which references snakes, mites, men, and divine power, among other things).

Tycho Brahe had developed a strong objection to the heliocentric theory of Nicolaus Copernicus. Johannes Kepler set out to answer that objection in his 1606 book *De Stella Nova*, his book on the "new star" of 1604 (now known to have been a supernova).

Brahe's objection was rooted in the stars. In the heliocentric theory the stars had to be very distant in order to explain why Earth's annual motion around the Sun produced no corresponding visible annual changes in their appearance—no "annual parallax". For instance, stars were not seen to grow brighter when Earth happened to move toward them as it journeyed around the Sun, nor were they seen to grow dimmer when it moved away from them. The explanation for this was that the orbit of the Earth was like a point in comparison to the distance to the stars—negligible in size.

But Brahe noted that stars have a measurable apparent size as seen from Earth. He had measured these sizes. He determined that the more prominent or "first magnitude" stars measured a little less than a tenth the apparent diameter of the Moon—a little less than three minutes of arc, since the Moon has an apparent diameter of approximately thirty minutes, or one half of one degree. At the vast distances required for the stars in the heliocentric hypothesis, these apparent sizes translated into enormous physical sizes. Were Copernicus correct, every one of the stars would have to dwarf the Sun. The Sun would be a unique, small body in a universe of giants.[1]

A decade after Brahe died, Johann Georg Locher and his mentor Christoph Scheiner would neatly summarize Brahe's objection in their 1614 book *Disquisitiones Mathematicae*. They wrote that in the Copernican hypothesis the Earth's orbit is like a point within the universe of stars; but the stars, having measurable sizes, are larger than points; therefore, in the Copernican hypothesis every star must be larger than Earth's orbit, and of course vastly larger than the Sun itself.[2]

The giant stars of the Copernican hypothesis stood in contrast to the more commensurate star sizes found in Brahe's own hypothesis, a hybrid geocentric (or geo-heliocentric) hypothesis in which the Sun, Moon, and stars circled an immobile Earth, but the planets circled the Sun (Figure 1). Brahe's hypothesis was observationally and mathematically identical to the Copernican hypothesis insofar as the Sun, Moon, and planets were concerned. However, since the Earth did not move relative to the stars in Brahe's geocentric hypothesis, there was no expectation of annual parallax, and thus no need for the stars to be distant in order to explain the absence of observable parallax. Brahe had the stars located a bit beyond Saturn. And, since the stars were roughly similar to Saturn in both distance and in their appearance in the night sky, they had to be similar to Saturn in physical size, too. In Brahe's hypothesis, the sizes of the Earth, Sun, Moon, and planets were commensurate, with the Moon being smallest and the Sun being largest, as opposed to the case in the Copernican hypothesis, where every last star dwarfed Sun, Moon, and planets (see Figure 2).[3]

---

[1] (Graney 2015, 32-38)

[2] (Graney 2017, 30)

[3] (Graney 2015, 32-38)

Kepler devotes Chapter 16 of *De Stella Nova* to the star size issue. This follows a discussion of the vast distance of the *stella nova* from Earth in Chapter 15. A translation of Chapter 16 into English is provided here in Appendix 1. The historian of science Albert van Helden has written that Brahe's measurements of the apparent sizes of stars were unassailable, and his logic incontrovertible, so Copernicans simply had to accept Brahe's objection and agree that the stars were giant.[4] This Kepler does. He does not question that the stars are all vastly larger than the Sun. In fact, in Chapter 16 he grants that a star with an apparent diameter of three minutes, one tenth the diameter of the Moon, has the same physical size as the orbit of Saturn; and that Sirius, the most brilliant of stars, is even larger; and the *nova* larger still. It follows from Kepler's numbers that any star whose physical size was the same as Earth's orbit would have an apparent diameter of three tenths of a minute, or eighteen seconds, the apparent diameter that Brahe had determined for the stars barely visible to the eye (sixth-magnitude stars).[5] And as the physical diameter of the Sun is less than one hundredth that of Earth's orbit according to Kepler, clearly every last star in the sky utterly dwarfs the Sun. To Kepler, the Sun and its planets are surrounded by giants, and only by giants.

Kepler's answer to Brahe then is that the Sun and planets being surrounded by giants makes sense, or at least more sense than the geocentric alternative. Kepler argues that Brahe fixes upon size, but that what is commensurate in the Copernican hypothesis are speeds. Speeds in any geocentric hypothesis are more incommensurate than sizes in the heliocentric hypothesis, he says. Moreover, he says, a vast range of sizes exists in the physical world, and he illustrates some of these: the longest snake vs. the smallest insect; human beings vs. the Earth; the Earth vs. the universe. And finally, there is the power and creativity of God, for whom nothing is too big, and yet who also confers value upon the small.

Answers such as these to Brahe's star size objection to Copernicus would endure. In 1651 Giovanni Battista Riccioli in his *Almagestum Novum* analyzed one hundred and twenty six pro- and anti-Copernican arguments, concluding that the vast majority in either direction were indecisive. As he saw it, there were two decisive arguments, both in favor of the anti-Copernicans: one was the absence of any detectable Coriolis Effect (as it would be called today);[6] the other was Brahe's star size objection. By 1651 the telescope had long been brought to bear on the star size question, including by Riccioli,[7] but since the small-aperture telescopes of the time showed stars as definite disks (not yet understood to be spurious artefacts of diffraction; see Figure 3) but did not reveal any parallax, Brahe's objection still stood (Simon Marius in his 1614 *Mundus Jovialis* first noted this telescopic support for Brahe[8]). Riccioli noted how Brahe's objection could be answered by appealing to the speed issue, but he dismissed this answer. The rising and setting of the stars is caused by either the rotation of the Earth or the rotation of the stars, he said, and in either case, whatever rotates turns though one circumference per day— proportionally the rates of motion are the same. And as for appealing to the power of God, that answer cannot be refuted, said the Jesuit Riccioli, but it does not satisfy the prudent. Besides, he said, if divine power can be called in as an explanation for the difficult aspects of a hypothesis, could not the geocentric hypothesis's vast speeds also be explained via divine power?[9]

---

[4] (Van Helden 1985, 51)

[5] (Graney 2015, 34) Brahe reported sixth-magnitude stars as having a diameter of twenty seconds.

[6] (Graney 2015, 115-128), (Graney 2017, xix)

[7] (Graney 2015, 53-61, 129-139)

[8] (Graney 2015, 50-53)

Thus whereas Johannes Kepler set out in *De Stella Nova* to answer Tycho Brahe's star size objection to the heliocentric theory of Nicolaus Copernicus, that objection still carried force almost five decades later. What would answer Brahe's objection would not be comparisons to geocentric speeds, or discussions of the sizes of snakes and mites, or appeals to divine power. Rather it would be the discovery that the apparent sizes of stars, whether measured visually or with a telescope, were the spurious product of optical systems, a product which gave no indication of the true sizes of stars. The first evidence suggesting the spurious nature of apparent stellar sizes, Jeremiah Horrocks' observations that stars winked out instantaneously when being occulted by the Moon, was not published until a decade after Riccioli's *Almagestum Novum*, six decades after *De Stella Nova*.[10] Such evidence would eventually show that stars did not have to all be giants in a Copernican universe. Indeed, recent progress in astronomy has shown that, while some giant stars do exist that dwarf the sun, these are relatively rare; most stars are smaller than the Sun, with a large majority of stars being small, dim "red dwarfs" that are far outclassed by the Sun.

---

[9] (Graney 2015, 136-138)

[10] (Graney 2015, 150-151)

APPENDIX 1—A translation into English of Chapter 16 of Johannes Kepler's *De Stella Nova*. Translation by C. M. Graney. Kepler's original paragraph breaks are indicated by spaces. The translator has broken Kepler's text into additional paragraphs, indicated by indentations.

# CHAPTER 16[11]
# Concerning the immensity of the sphere of the fixed stars in the hypothesis of Copernicus: and concerning the size of the new star.

At this point I am sure, among those who are reading, are some who are ready to ridicule the insanity of Copernicus. Other readers, who embrace his opinion on account of different astronomical and physical arguments, are nevertheless troubled by this immensity, and are beginning to doubt whether those arguments (which are produced out of other knowledge for the purpose of establishing the opinion of Copernicus) may be asserted altogether truly, and whether the occasion may be right for those arguments to be dissolved by some other reasoning, and for it to be shown, how they (the readers) may be the dupes of Copernican error.

    Brahe has added force to this objection. He finds a lack of elegance in the most perfect of works, if the vastness of the sphere of one of the fixed stars be so insane; the meagerness of all the wandering stars so contemptible. How huge the fault in the human body, he says, if the finger, if the nose, might surpass by many times the bulk of the whole remainder of the body.
    Therefore, in order perchance to dilute this objection, I shall say three things. First, I shall show how many things more incredible come forth from the opinion of the ancients and of Ptolemy, concerning the movement of the fixed stars. Next, I shall establish, through different examples chosen from the world, the existence of great ranges of sizes and proportions. And lastly, I shall point out a mutual compensation by which some equality arises in the most unequal of things.

    Accordingly, the motion of the bodies of the universe in the hypotheses of Copernicus is such that in one hour the globe of Saturn swims across about 300

---

[11] (Kepler 1606, 83-89). Copies of Kepler's original book are freely available via Google Books and Erara. For secondary sources that discuss Chapter 16 see (Westman 2011, 398-399) and (Boner 2011, 101-106). Neither of these discuss at any length Kepler's attention to the issue of the sizes of the fixed stars themselves. Rather both focus on the question of the size of the universe as a whole. Boner does include a quotation from Kepler that remarks upon the size of the *stella nova* under the Copernican hypothesis as exceeding that of the planetary system (Boner 2011, 104).

German miles; Jupiter, 400; Mars, 600; the center of Earth, 740;[12] Venus, 800; Mercury, 1000—a beautiful proportion, where what is nearer to the quiescent Sun (the dispenser of all movement) is always swifter.  And because the Earth not only is conveyed annually; but also is revolved daily, in one hour the regions of the Earth which are near to the equator traverse 240 German miles around the center,[13] the rest of the regions less (as they are nearer to the poles).  Therefore the 240 miles of the surface added to the 740 of the center, make about 1000.  This is how much aethereal space some city situated at the equator traverses in one hour of the middle of the night.  And 240 subtracted from 740 leaves 500 miles remaining—the journey through the aethereal vapor of one midday hour at the equator.[14]

Whence it should be considered, because the nighttime speed of the lands at the equator is twice that of the daytime, whether this somewhat aids in cooling the nights, just like a kind of ventilation, and whether indeed on account of this, the morning time receives the greater favor.  For the parts of the Earth, on which the Sun now rises, at this moment are carried directly into the aethereal vapor.  Meanwhile the parts where the Sun sets are, at this time, as it were, pulled away from the aethereal vapor.

The Moon remains for us to consider.  In one hour it traverses 476 miles around the center of the Earth, because it is distant from Earth by 60 terrestrial semidiameters.[15]  Add 476 to the journey of the center of Earth, 740: the journey of the full Moon comes to 1216 miles through the aethereal vapor.  Subtract in turn 476 from the journey of the center, 740: the journey of the new Moon in one hour leaves behind 264 miles.

From Copernicus you have commensurate speeds of all the stars, fixed and wandering.  Indeed the fixed stars are like a place.  They are quiescent.  The immobile Sun is like a mover.  It maintains the middle station of the universe.  The body of the Sun, being rotated in place, circularly stirs by this rotation, throughout the fullness of the universe, the immaterial species (unless I merely concocted such a fitting circumstance by physical reasoning, in the commentaries concerning the movements of the star Mars), a mover which the wandering stars all follow. And what is nearer to the Sun is swifter.

---

[12] Kepler gives the Earth-Sun distance (1 astronomical unit, or 1 AU) as 1200 terrestrial semidiameters, which in turn he gives as 860 German miles (he states that a German mile is 5000 paces of 5 feet per pace).  Thus the circumference of Earth's orbit is $2 \times \pi \times (1200 \times 860) =$ 6,484,247 miles.  A year being 365.25 days and a day being 24 hours, this yields a speed of $6,484,247 / (365.25 \times 24) = 740$.  Speeds for other planets are calculated in the same way.  For example, Saturn, having an orbital radius roughly ten times that of Earth, and an orbital period roughly thirty times, has an orbital speed of roughly a third that of Earth.

[13] If the semidiameter of Earth is 860 German miles, the circumference of Earth at the equator is then $2 \times \pi \times 860 = 5404$ miles.  $5404 / 24$ hours $= 225$ rather than 240.

[14] Earth's orbital and rotational speeds add together on its the night side and subtract on its day side.  Galileo would use this concept in his discussion of the ocean tides being evidence for Earth's motion.

[15] The Moon takes 27.3 days to circle the Earth: $2 \times \pi \times 60 \times 860$ miles $/ (27.3 \times 24$ hours$) = 495$ rather than 476.

Lest this passage through a thousand miles in one hour still seem to you truly incredible, I urge you to consider the proportion of the density of air to the density of the aether, which I have demonstrated in Optics. There it is proven, how the passage of a thousand miles in one hour through the aether may be more tranquil, by an incredible number of times, than is the passage of one mile in one hour through our air.

Go now to Ptolemy and the ancient opinion; you will find everything more incredible. In that, the semidiameter of the sphere of the fixed stars occupies twenty thousand semidiameters of Earth. The circumference therefore will be sixty-three thousand[16]—truly a reasonable number, compared to the Copernican, but which all is said to go round in one day. Therefore 2625 semidiameters (each of which contains 860 miles) are covered in one hour.[17] Behold here what to me is an immense distinction. In the view of Ptolemy, Saturn is the nearest to the fixed stars, such that it will almost touch them. Following Copernicus, in one hour it traverses 300 miles; following Ptolemy, twenty two hundred thousand fifty seven thousand five hundred miles.[18] Saturn must be believed to be seven thousand, five hundred twenty five[19] times swifter under Ptolemy, than under Copernicus. Whoever attempts mentally to comprehend this incredible velocity is overcome just as much as, and indeed more severely than, someone who attempts to comprehend the Copernican immensity.

Meanwhile, weigh carefully, O Philosopher, the proportion of accident to its subject to be desired so much here, against who by right is able to desire proportion of part to part of the universe in the view of Copernicus. For accidents are not without suitable subject. More credible is a great subject without motion, than a great motion in a small subject. And so, if I have carried on regarding the Ptolemaic proportion of movements versus the Copernican proportion of bodies, I have done it to the greater applause of the Philosophers.

Ptolemy multiplies the motion of Aristarchus or Copernicus seven and one half thousand times. I shall multiply by that many times the Ptolemaic body of the fixed stars, 20,000 semidiameters, so they may rise to fifteen hundred thousand semidiameters, upped one hundred times. This is 150,500,000.[20] Dividing into this twelve hundred, which is the measure of the interval between the Sun and the Earth, yields one hundred twenty five thousand solar intervals—specifically 125,417.[21]  But we do not require that much bulk in order to defend Copernicus.

---

[16] $20,000 \times \pi = 63,000$. As circumference is $\pi$ times diameter, not semidiameter, this number is too small by half. There are a variety of such typos in Kepler's text of Chapter 16.

[17] $63,000 / 24 = 2625$

[18] $860 \times 2625 = 2,257,500$

[19] $2,257,500 / 300 = 7,525$

[20] $20,000 \times 7,525 = 150,500,000$

We can compliment the universe for its elegance of proportion—not for proportion of size, what Brahe has been in the habit to desire, but rather for proportion of beauty and reasoning.  The perfection of the universe is motion, which is as it were a certain life of it.  Towards motion three things are required: a mover, a movable, a place.  The mover is the Sun.  The movables are Mercury up through Saturn.  The place is the far sphere of the fixed stars.

But, if it is permitted to express the physical thing mathematically, the movables are the proportional middle between mover and place.  For even they aspire to the quiescence of the place or of the encircling body.  Resisting, as if by a certain heaviness (you ridicule what, O inexpert spinners of philosophical phrases?[22]—you rich in plenty of the imaginary things celestial, most destitute of the true), they receive motion from the mover—out of which their individual periods of time fall out, opposite to how close each holds to the mover (it being unitary, and moving uniformly).  Thus, in the movables, both the movement and the quiescence are in a certain way mingled.  But if the movables are physically the proportional middle between mover and place, what has more verisimilitude, even mathematically, than for the diameter of the movables to be the proportional middle between the diameter of the mover Sun, and the diameter of the place or of the fixed stars?

This posited, it is easy to investigate the quantity of the fixed stars.  For the diameter of the Sun surpasses the diameter of Earth by five and a half times.  It may surpass by fully six, because I say, I attribute not 1200 but 1432 semidiameters to the distance between the Sun and Earth.  Yet I may be being generous, in order that the universe may be made altogether large.

In the accompanying figure [Figure 4] the semidiameter CE of the body of the Sun is 6; the distance CK from the Sun to the Earth is 1432.  The more lofty sphere of Saturn is GB.  Now Saturn is ten times the Sun's distance—specifically, CG is ten times CK.  It is indeed somewhat less, but nevertheless this number will do.  Therefore Saturn is distant by 14,320 semidiameters of Earth; this is the length of CG.  And so now, as the semidiameter of the solar body CI[23] (six semidiameters of Earth) to the semidiameter of the orb of Saturn CG or CB (14,320 semidiameters of Earth); so this CB to the semidiameter of the fixed stars CD, formed by continuing CG to the fixed stars.[24]  Upon this line semicircle EBD is drawn, carried across through marks E and B, the center of which is A.

---

[21] 150,500,000 / 1200 = 125,417

[22] In the traditional Aristotelian view, celestial bodies were supposed to be made of an aetherial substance and would have no heaviness.

[23] Kepler's text reads CO, but this is certainly a typo.

[24] CI / CB = CB / CD

Reaching the fixed stars, CG will manifest to us the maximum number of semidiameters of Earth, three hundred forty hundred seventy seven thousand. This is 34,077,066⅔.[25] Yet such a maximum is not one quarter that previous number [150,500,000], which we had taken up according to the example of multiplied speed in the old opinion.

Therefore the philosophers fuss about what?—removing from the eye of Copernicus this mote of the immensity of the fixed stars, while at the same time ignoring in their own eye, the enormous log, more than four times greater, of the insane speed of the fixed stars.

Tycho Brahe has somewhat helped the old opinion here, reducing the loftiness by somewhat more than a third, because he puts that at seven thousand semidiameters of Earth for Saturn, and he truly lavishes the fixed stars with double this—this indeed from his own supposed model of the universe, by necessary demonstration, from truly plausible conjecture. But whether the one, or whether the two thirds; still both ways exceed, by three or four times.[26] And what is one star of Saturn to the monstrous multitude of the fixed stars?—which all in the view of the received hypothesis are moved by that most swift motion; which all in the view of Copernicus rest immobile, only seven little bodies marching along [5 planets, Earth, Moon].

Yet might it be that the universe, adorned by the proportion here fully pointed out by me, is still narrow, and so close that the fixed stars may make some parallax?—that is, that the proportion of line CK or CO, by which the Sun and Earth stand apart, may become sensible compared to CD, the separation of the Sun and the fixed stars? Not at all. For that previous number, namely three hundred forty hundreds of thousands [34,000,000], is twenty times the distance that makes a parallax of two minutes—in fact a little less, just over sixteen times. And so three hundred forty, etc. claims a parallax of a little less than a sixteenth of two, namely of approximately one eighth of a minute.[27]

---

[25] $(14320 / 6) \times 14320 = 34,177,067$

[26] Kepler's text says "four or five", but I take this as a typo since the math is "three or four".

[27] Kepler's words here regarding the parallax calculation are very difficult, and these two English sentences are a product of interpretation as much as of translation. Kepler seems to start with the distance that will produce a parallax of two arc seconds, or 2/60 of a degree—*distantia... potuit facere duorum minutorum parallaxin*. Tan(2/60)=0.0005236, and the inverse of that, 1719, is the distance to the stellar sphere (CD in Kepler's figure), measured in AU's, that will produce a two minute parallax. Sixteen times this value is 27,502 AU. If we suppose that Kepler used his earlier value of 1200 semidiameters of the Earth as his AU value for this calculation, then we get $1200 \times 27,502 = 33,000,000$. As Kepler specifies that the multiplier is a little more than sixteen, specifying a parallax of approximately an eighth of a minute as a little less than a sixteenth of two-minutes—*sedecimum hujus paulo minus*—we can take the multiplier as sixteen and one-half, and we get $1719 \times 16.5 \times 1200 = 34,000,000$. Happily, the sentences that follow provide a much clearer discussion regarding parallax.

Or, in another way, more artfully and briefly: Because the Sun occupies 30 minutes, seen from Earth, that is, because angle CKI is 15′, and certainly Saturn is higher by tenfold, then the Sun, seen from Saturn, occupies a tenth part of 30 minutes—that is 3 minutes. Truly this quantity is double the angle CGI. But it is posited, as IC the Sun to CG the semidiameter of the circle of Saturn, thus CG or CB to CD the semidiameter of the fixed stars. And so, connecting points BD, triangles BCD and ICG will be similar; hence angles BDC[28] and IGC equal. Therefore even that circle of Saturn CB, seen from the sphere of the fixed stars, or point D, might produce an appearance of 1½ minutes. And since CO the circle of Earth[29] may be the tenth part of the semidiameter[30] of Saturn CB, therefore angle ODC will be about the tenth part of angle BDC. Thus the semidiameter of the annual orb of Earth CO is not more than 9″,[31] so it will cause a parallax of the seventh part of one minute.[32]

But no astronomer has the observational skill needed to be able to boast of achieving an observation of a seventh or eighth part of a minute. For instance, though the separation of two stars near mark D may be less by the seventh part of a minute when the Earth is moving through O (when D is at a right angle to the Sun or to line OC), than when the Earth is moving through K (when D is opposite of the Sun), the astronomer still will not distinguish these distances, observing them both at O quadrature and K opposition of the Sun, since the bodies themselves of the fixed stars generally occupy and subtend one, two, three, or four minutes.

So therefore it is shown that no one who embraces whatever is true from the hypotheses, and who does not wish the whole of Astronomy utterly overturned, is prudent to present the immensity of the universe as an objection to Copernicus.

Now it still needs to be seen, by just what examples that immense proportion may become palatable to us. Here I have usually presented to Brahe as an objection the proportion of the mite which burrows into the skin of the hand to that 120 foot[33] serpent mentioned by Pliny. Hides of it were preserved at Rome, and Brahe was declaring that larger ones had been seen in the North. The proportion of length was extending to a hundred thousand.[34]

---

[28] Original reads BCD. In the following sentence the original reads BDC.

[29] *telluris coelum*

[30] Original reads *diametri*.

[31] 1.5′ / 10 = 0.15′ = 0.15 × 60″ = 9″

[32] 60/7 = 8.57 ≈ 9

[33] Kepler states this value in Roman numerals: CXX.

[34] 120 feet / 100,000 = 0.0012 feet, which is roughly half of a millimeter—a reasonable estimate for a barely-visible mite.

But this example is a stunt.  Take another.  I ask, how small is a man compared to the globe of Earth?  We may compute.  From the surface to the center are 860 miles, each of which reckons into 5000 paces.  Therefore there are 4,300,000 paces.[35]  Five times that is feet, truly 21,500,000.[36]  Grant to the length of a man seven whole feet, and divide that number through those.  Therefore if you compare length, thirty one hundred thousand men[37] in succession will reach from the surface of Earth to the center of it.  And one diameter of Earth equals more than six hundred myriads[38] of human height.  All confess this immense proportion of man to globe.  But now they call incredible the proportion of Earth to the circle[39] of Saturn, which is one to twelve thousand.[40]  They call incredible the proportion of the realm of the movables[41] to the immobile sphere of the fixed stars,[42] which is one to three thousand.[43]  These both are much less than that.  Certainly they do not consider proportion, but magnitude, even though they may be small.  Truly the universe is not large to God, but we are small to the universe.

Yet consider again an analogy.  Where magnitude waxes, there perfection wanes, and nobility follows diminution in bulk.  The sphere of the fixed stars according to Copernicus is certainly most large; but it is inert, no motion.

The universe of the movables is next.  Now this—so much smaller, so much more divine—has accepted that so admirable, so well-ordered motion.  Nevertheless, that place neither contains animating faculty, nor does it reason, nor does it run about.  It goes, provided that it is moved.  It has not developed, but it retains that impressed to it from the beginning.  What it is not, it will never be.  What it is, is not made by it—the same endures, as was built.

Then comes this our little ball, the little cottage of us all, which we call the Earth: the womb of the growing, herself fashioned by a certain internal faculty.  The architect of marvelous work, she kindles daily so many little living things from herself—plants, fishes, insects—as she easily may scorn the rest of the bulk in view of this her nobility.

Lastly behold if you will the little bodies which we call the animals.  What smaller than these is able to be imagined in comparison to the universe?  But there now behold feeling, and voluntary motions—an infinite architecture of bodies.

Behold if you will, among those, these fine bits of dust, which are called Men; to whom the Creator has granted such, that in a certain way they may beget themselves, clothe themselves, arm themselves, teach themselves an infinity of arts, and daily accomplish the good; in whom is the image of God; who are, in a certain way, lords of the whole bulk.

---

[35] $860 \times 5000 = 4,300,000$

[36] $4,300,000 \times 5 = 21,500,000$

[37] $21,500,000 / 7 = 3,071,429$

[38] One definition of a "myriad" is ten thousand.

[39] *coelum*—the "heaven" or celestial sphere in which the body of Saturn travels.

[40] Although earlier Kepler stated 1 AU to be 1432 terrestrial semidiameters rather than 1200, here he uses 1200: that multiplied by Saturn's ten-fold greater distance from the Sun yields 12,000.

[41] *coelorum mobilium*

[42] *coelum fixarum immobile*

[43] $34,177,067 / 12,000 = 2848 \approx 3000$

And what is it to us, that the body of the universe has for itself a great breadth, while the soul lacks for one?  We may learn well therefore the pleasure of the Creator, who is author both of the roughness of the large masses, and of the perfection of the smalls.  Yet he glories not in bulk, but ennobles those which he has wished to be small.

In the end, through these intervals from Earth to the Sun, from Sun to Saturn, from Saturn to the fixed stars, we may learn gradually to ascend toward recognizing the immensity of divine power.

I have gladly inserted so much here concerning the objections to the Copernican vastness of the fixed stars, because it all pertains to the incredible magnitude that must be estimated for the new star.  For if it occupies only four minutes (the size Sirius appears), then through this hypothesis of Copernicus it is much greater than the whole machinery of the movables.  For earlier we were granting to that machinery only three minutes, were it to be seen from the fixed stars.  Indeed I refrain to express such magnitude by numbers, dreading that I may have already thrust forward those objections excessively, to the point of generating derision by the unlearned commoner.

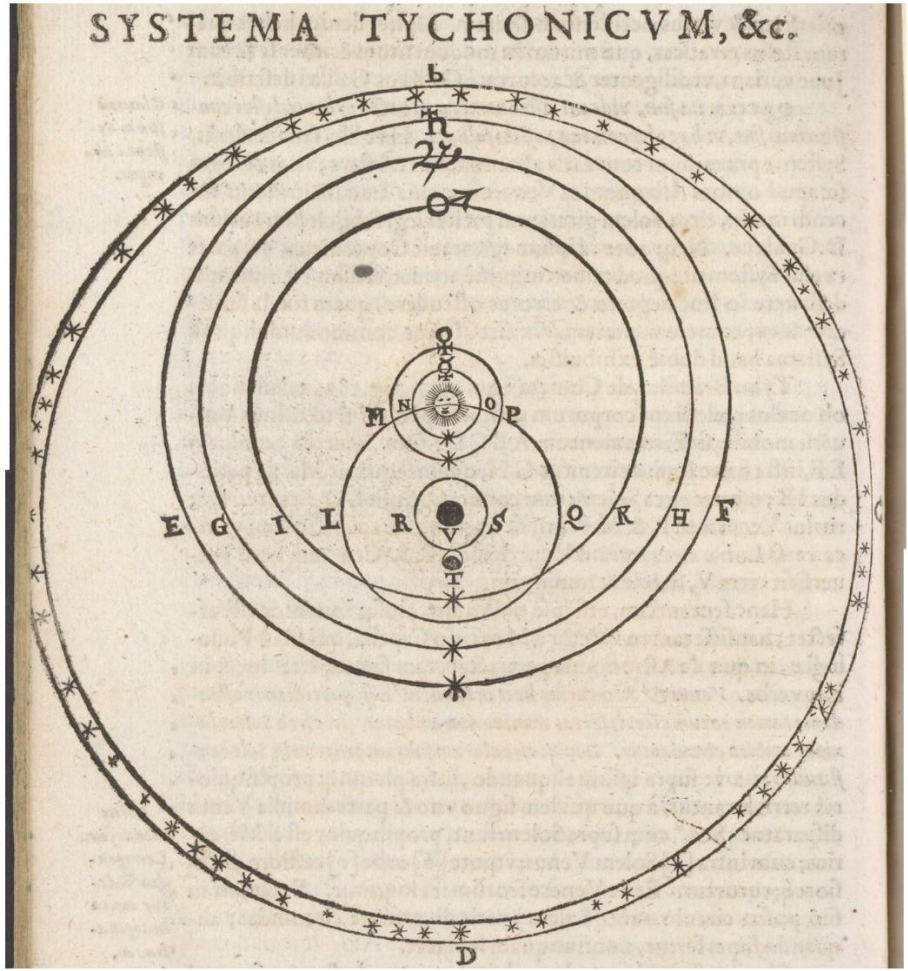

FIGURE 1. Tycho Brahe's hypothesis. Earth is immobile at center. Mercury, Venus, Mars, Jupiter, and Saturn circle the Sun as in the Copernican hypothesis, while the Sun circles the Earth (as do the Moon and stars). From (Locher 1614, 52). Image credit: ETH-Bibliothek Zürich, Alte und Seltene Drucke.

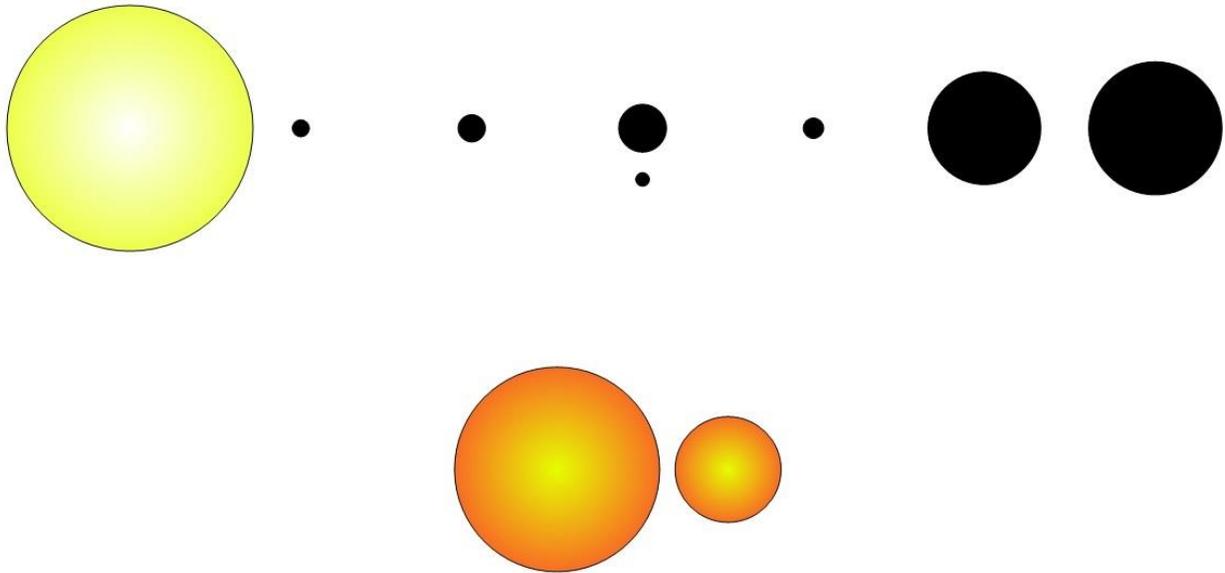

FIGURE 2. *Above*—the relative sizes of celestial bodies calculated by Tycho Brahe, based on his observations and measurements, for (from left to right, upper row) the Sun, Mercury, Venus, Earth and Moon, Mars, Jupiter, Saturn, as well as for (lower row) a large star and a mid-sized star in a hybrid geocentric universe (where the stars lie just beyond Saturn, as in Figure 1). Sun, stars, and planets all fall into a fairly consistent range of sizes. *Below*—the arrowed dots are the figure above, reproduced to scale compared to Brahe's calculated relative size for a mid-size star in the Copernican universe (where the stars lie at vast distances, and thus must be enormous to explain their apparent sizes as seen from Earth). Brahe said the huge Copernican stars were absurd.

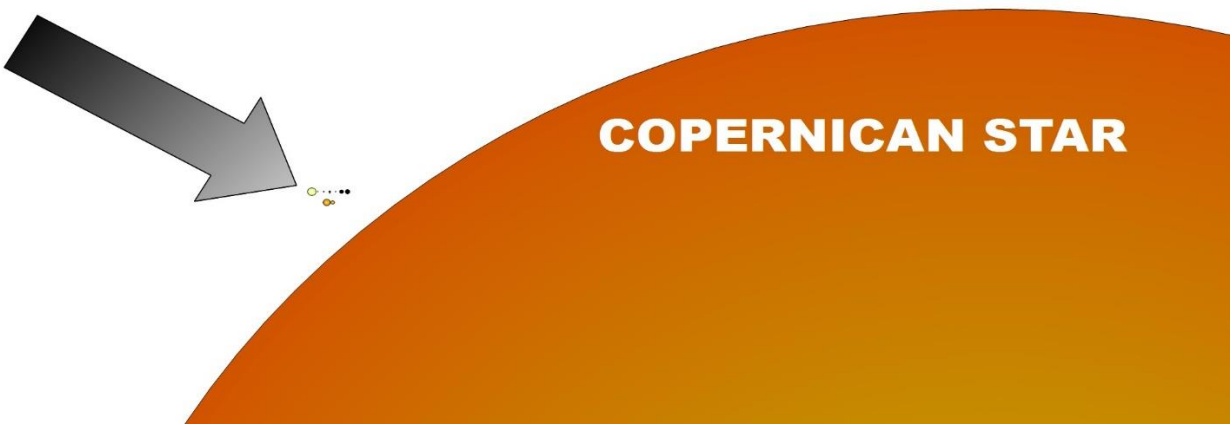

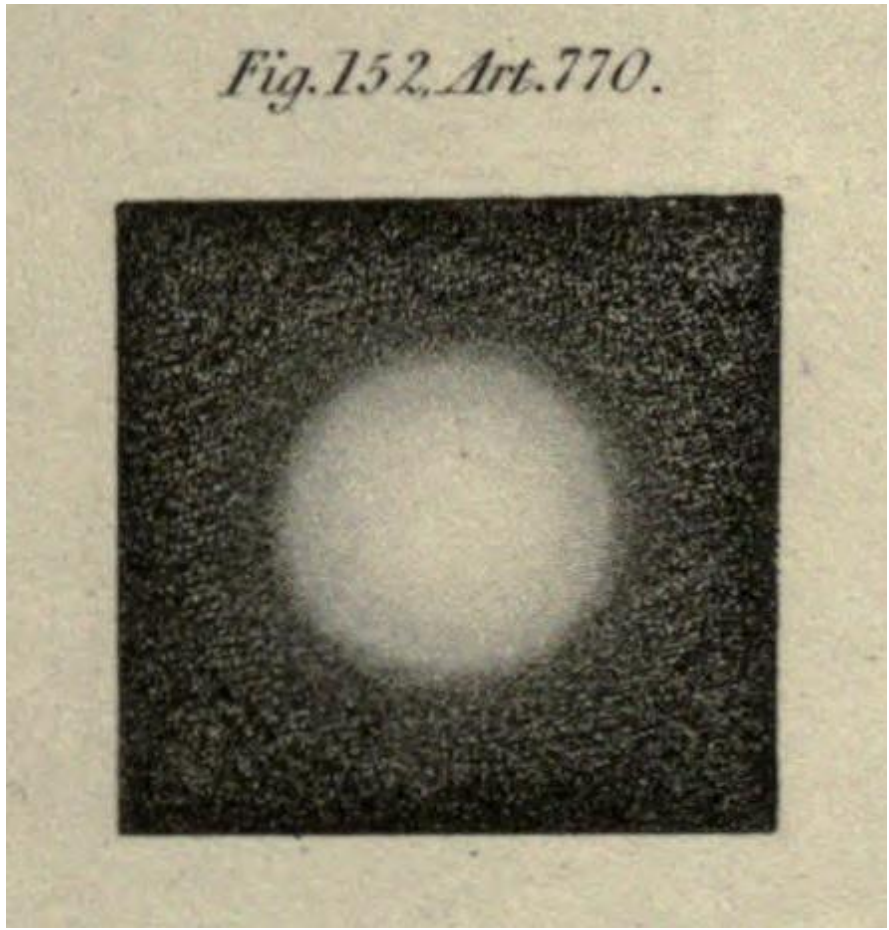

FIGURE 3. A star as seen through a small aperture telescope (Herschel 1828, 491 & Plate 9). This appearance of a sphere of measurable size is entirely spurious—an artifact of diffraction. However, early telescopic astronomers took such telescopic images to be the physical bodies of stars (Graney and Grayson 2011). Image credit: ETH-Bibliothek Zürich, Alte und Seltene Drucke.

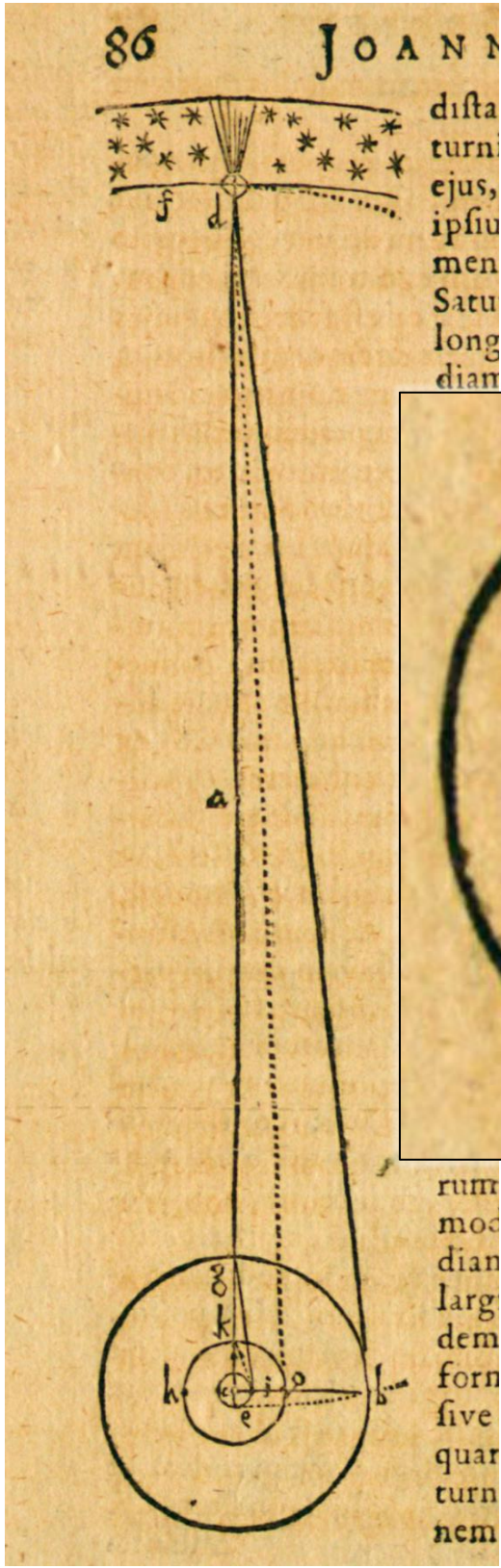
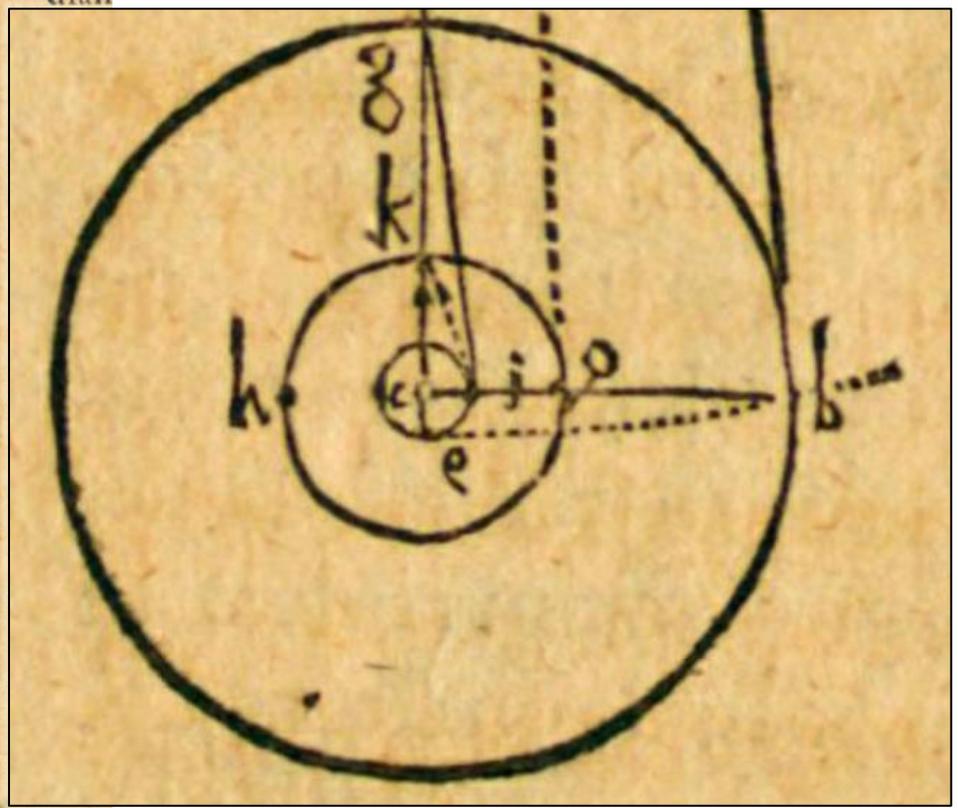

FIGURE 4. Kepler's illustration showing the globe of the Sun, the orbit of Earth, the orbit of Saturn, and the sphere of the fixed stars. Image credit: ETH-Bibliothek Zürich, Alte und Seltene Drucke.